\documentclass[aps, showpacs, pra, superscriptadaddress, 
twocolumn] {revtex4-1}
\usepackage{hyperref}
\usepackage{amsmath }
\usepackage{amssymb}
\usepackage{epsfig}
\usepackage{natbib}
\newcommand*{\be}{\begin{equation}}
\newcommand*{\ee}{\end{equation}}
\newcommand*{\bea}{\begin{eqnarray}}
\newcommand*{\eea}{\end{eqnarray}}
 
 \DeclareFontFamily{OT1}{pzc}{}
 \DeclareFontShape{OT1}{pzc}{m}{it}%
 {<->  s  *  [1.400]  pzcmi7t}{}
\DeclareMathAlphabet{\mathscr}{OT1}{pzc}%
{m}{it}

\begin{document}

\title[]{$\mathcal{PT}$-symmetry breaking in a necklace of coupled optical waveguides}

\author{I  V  Barashenkov$^{1,2}$,  L  Baker$^1$, and N V Alexeeva$^{1,2}$}
 \affiliation{$^1$
 Department of Mathematics and Centre for Theoretical  and Mathematical Physics,
 University of Cape Town, Rondebosch 7701, South Africa \\
 $^2$  
 New Zealand Institute for Advanced Study, Massey University, Auckland 0745, New Zealand
 }

\begin{abstract}
We consider parity-time ($\mathcal{PT}$) symmetric arrays
   formed by  
 $N$ optical waveguides with  gain and $N$ waveguides with loss.
When the gain-loss coefficient exceeds a critical value $\gamma_c$, the $\mathcal{PT}$-symmetry
becomes spontaneously broken. We  calculate $\gamma_c(N)$ and prove that  $\gamma_c \to 0$
as $N \to \infty$. In the symmetric phase, the periodic array is shown to support $2N$ solitons with different
frequencies and  polarisations.
\end{abstract}

\maketitle

\section{Introduction}

Since the emergence of the $\mathcal{PT}$  symmetry 
as a research avenue in quantum theory \cite{BB}, the concept was embraced in several other fields,
including photonics \cite{Rueter}, plasmonics \cite{plasmonics}, Bose-Einstein condensates \cite{Graefe,BEC}, and quantum optics of atomic gases \cite{Lambda}.
 The $\mathcal{PT}$-symmetric systems exhibit unusual  phenomenology
with a potential for practical utilisation. 
The $\mathcal{PT}$  optical structures, in particular, 
 display 
unconventional beam refraction \cite{Musslimani,Zheng}, Bragg scattering \cite{Berry_Longhi},
nonreciprocal Bloch oscillations \cite{Longhi}, 
  loss-induced  transparency \cite{Guo}, and
conical diffraction \cite{Ramezani}.  Nonlinear effects in such systems 
can be utilised  for an efficient control of light, including
all-optical low-threshold switching \cite{RKEC,SXK}
and unidirectional
invisibility \cite{RKEC,Lin}.

Experimentally,  the optical $\mathcal{PT}$ symmetry  was realised in a directional coupler 
consisting of two
coupled  waveguides with gain and loss \cite{Guo,Rueter},
and in chains of such dipoles \cite{Regensburger}. 
The corresponding theoretical models  went on to include the effects of  diffraction of
spatial beams and dispersion of  temporal pulses,  i.e., include an additional spatial or temporal dimension
\cite{couplers}. Dispersive $\mathcal{PT}$-couplers were shown to support optical solitons  \cite{couplers,breather}. 
Triplets, quadruplets and quintets of (nondispersive) guides were also dealt with \cite{oligomers}.

A fundamental phenomenon observed in symmetric couplers is the spontaneous $\mathcal{PT}$-symmetry
breaking \cite{Klaiman,Lin,Rueter,Regensburger,couplers}
which occurs as the gain-loss coefficient is increased beyond a certain critical value $\gamma_c$.
This exceptional point separates the symmetric phase, where all perturbation frequencies are real,
and the symmetry-broken phase, where some frequencies are complex and the corresponding 
modes grow exponentially. 
Besides demarcating the stability boundary, the exceptional point has several other roles to play.
In particular, it is in the vicinity of this critical value that  
the $\mathcal{PT}$-symmetric periodic structures
can act as unidirectional invisible media \cite{Lin}.

\begin{figure}[t]
 \begin{center}
      \includegraphics*[height=35mm, width=0.3\linewidth]{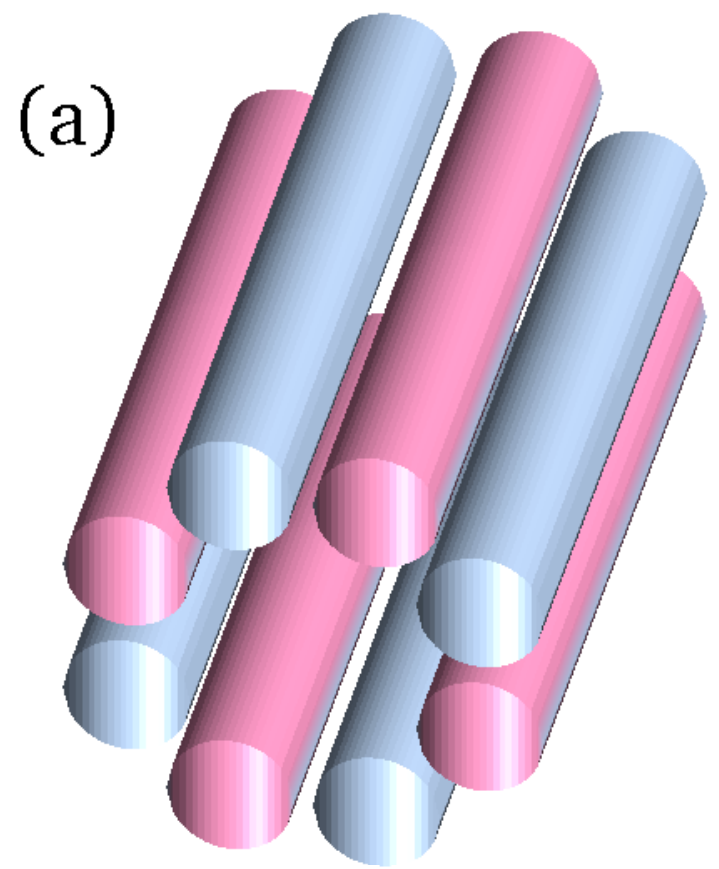}
     \hspace*{0.15\linewidth}
      \includegraphics*[height=35mm, width=0.3\linewidth]{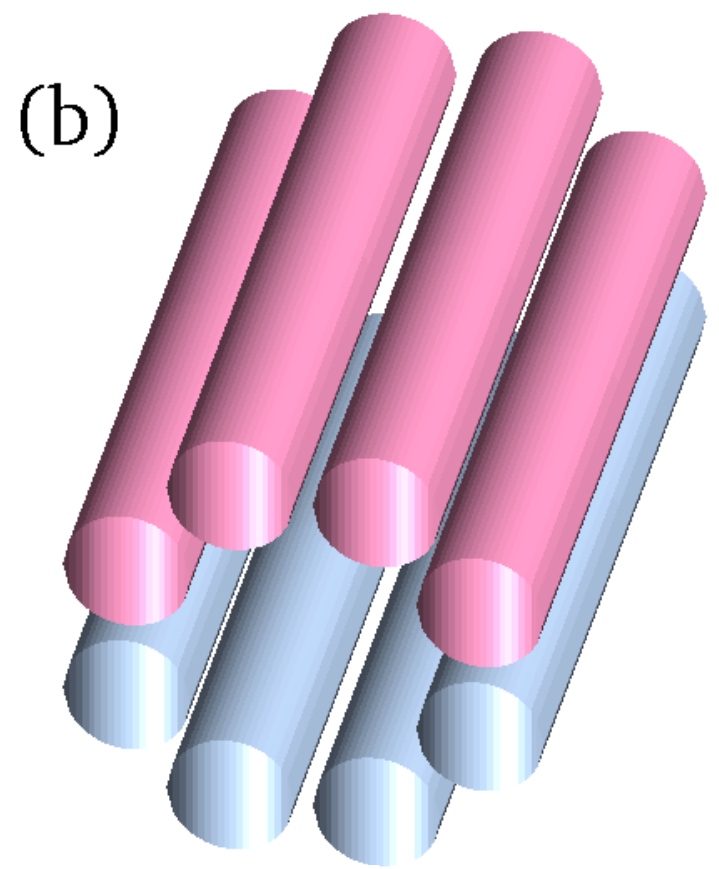}           
             \caption{(Color) An alternating (a) and clustered (b) necklace of waveguides
                with gain (pink) and  loss (blue).
   \label{fig1}}
 \end{center}
 \end{figure}

In this paper, we consider a  generalisation of the two-channel
dispersive coupler to a $\mathcal{PT}$-symmetric arrangement  of $2N$ dispersive waveguides  coupled  to their nearest neighbours
--- see fig.\ref{fig1}. (The previously analysed situation corresponds to $N=1$ \cite{couplers,breather}.)
  The issue that concerns us here, is how the geometry of this system and the growing multiplicity of its channels affects the symmetry-breaking point, $\gamma_c$. 
We also uncover  the 
 diversity of solitons arising in such a chain.

The chain consists of $N$ waveguides with gain and $N$ with loss.
The complex mode  amplitudes, $u_n$,  satisfy
\begin{align}
i {\dot u_n} + u_n^{\prime \prime}+ 2|u_n|^2 u_n +u_{n-1} + u_{n+1}= 2 i \Gamma_n u_n \nonumber  \\
(n=1,...,2N),
\label{Y1}
\end{align}
where ${\dot u_n} \equiv \partial u_n/\partial t$ and
 $u_n^{\prime \prime} \equiv \partial^2 u_n/\partial z^2$. 
 In Eq.\eqref{Y1},  $t$ stands for time and $z$  for the  distance in the
frame of reference traveling along with the pulse.
   The  coefficient  $\Gamma_n$ equals $\gamma>0$ for the waveguides with gain and 
 $-\gamma$ for those with loss. The active and lossy guides are either
 separated into two clusters or simply alternate (fig.\ref{fig1}). 
The chain  forms a periodic necklace,
that is,  
$u_{2N+1}=u_1$
and $u_0=u_{2N}$. 

We also consider open chains. 
An open chain is described by Eqs.\eqref{Y1} without the periodicity condition:
\begin{align}
i {\dot u_1} + u_1^{\prime \prime}+ 2|u_1|^2 u_1 +u_2 = 2i \gamma u_1,  
\nonumber 
\\
i {\dot u_{2N}} + u_{2N}^{\prime \prime}+ 2|u_{2N}|^2 u_{2N} +u_{2N-1} = -2i \gamma u_{2N},
\nonumber  \\
i {\dot u_n} + u_n^{\prime \prime}+ 2|u_n|^2 u_n +u_{n-1} + u_{n+1}= 2 i \Gamma_n u_n
\nonumber \\
(n=2, ... , 2N-1).
\label{Y2}
\end{align}
  With a suitably chosen constant matrix $\mathcal{L}$,  Eqs.\eqref{Y1}-\eqref{Y2} 
 can be written in a unified way:
 \be
 i {\dot u_n} + u_n^{\prime \prime}+ \sum_{m=1}^{2N}  
 \mathcal{L}_{nm} u_m  + 2|u_n|^2 u_n =0.
 \label{Y100}
 \ee

Of particular importance for dispersive waveguides is the zero solution of Eq.\eqref{Y100},
$u_n(z,t)=0$ ($n=1,...,2N$). This solution shall serve as a background to solitons  \cite{couplers}
and breathers \cite{breather}.
  To classify its stability, we 
linearise  \eqref{Y100} and let
$u_n= c_n e^{i(kz-\omega t)}$, where $k$ and $\omega$ are
assumed to be real. 
The combination 
$\lambda= k^2-  \omega$
is then found as an eigenvalue of the linearisation matrix $\mathcal{L}$.
The zero solution loses stability when two real eigenvalues merge and
become a complex-conjugate pair.

It is fitting to note that an equivalent eigenvalue problem arises in the linearisation of
the trivial solution of the symmetric array of {\it nondispersive\/} ($z$-independent) waveguides.
[The nondispersive array is a lattice system defined by 
Eq.\eqref{Y100} without the $u_n^{\prime \prime}$ term.]
The $N=1$ case corresponds to the $\mathcal{PT}$-symmetric nondispersive coupler
(also referred to as the dimer) \cite{Graefe,RKEC,SXK}.
 The  $N=2$ case (the $\mathcal{PT}$-quadrimer) was considered in \cite{oligomers}.

\section{Open alternating chain}

The alternating necklaces have $\Gamma_n= (-1)^{n+1} \gamma$ for $n=1, ..., 2N$. 
It is convenient to consider the {\it open\/}  chain first.
In this case,
\be
\mathcal{L}_{nm}= - 2 i \Gamma_n
\delta_{n-m} + \delta_{n-m -1}+ \delta_{n-m+1}, 
\label{Y11}
\ee
$n,m=1, ..., 2N$. Here $\delta$ is the Kronecker delta symbol: 
\[
\delta_n= \left\{ 
\begin{array}{ll} 
1, & \mbox{if} \ n= 0; \\
0  &  \mbox{otherwise}.
\end{array}
\right.
\]

The stability eigenvalues are expressible via roots of the secular equation
$\mathcal{D}_{2N}(\gamma, \alpha)=0$,
where $\lambda=-2\alpha$ and
\be
\mathcal{D}_{2N}= \mathrm{det} \, (\mathcal{L}+ 2\alpha I).
\label{D}
 \ee
The determinants with $N=1$ and 2 are readily found:  
\begin{align*}
\mathcal{D}_2=2x+1; \\
\mathcal{D}_4=4x^2+2x-1, 
\end{align*}
where $x=2(\gamma^2+\alpha^2)-1$. 
Any determinant with $N \geq 3$  can
be expanded as
\be
\mathcal{D}_{2N}= 2 x {\mathcal D}_{2(N-1)} - \mathcal{D}_{2(N-2)}.
\label{C1}
\ee
The recursion relation \eqref{C1} with 
$\mathcal{D}_{2}=2x+1$ and
$\mathcal{D}_4=4x^2+2x-1$ admits a simple  solution
\be
\mathcal{D}_{2N}(\gamma, \alpha)= U_N(x) + U_{N-1} (x),
\label{C2}
\ee
where $U_N(x)$ 
is the Chebyshev polynomial
of the second kind of $N$th order \cite{Cheb}.
Using the defining property of the  Chebyshev polynomials,
\[
U_N(x)= \frac{\sin[(N+1) \theta]}{\sin \theta},  \quad
\mbox{where} \  x= \cos \theta,
\]
we evaluate
the determinant $\mathcal{D}_{2N}$ as
\be
\mathcal{D}_{2N}= \frac{\sin[(2N+1)\theta/2]}
{\sin(\theta/2)}.   \label{Y4}
\ee
$\mathcal{D}_{2N}$ has $2N$ simple roots 
\[
\theta_n= \pm \frac{2n}{2N+1} \pi, \quad 
n=1,2, ..., N.
\]

On the $(\gamma, \alpha)$-plane, equations $\gamma^2 +\alpha^2 = \cos^2 (\theta_n/2)$ describe 
  $N$ concentric circles  centred at the origin.   As $\gamma$ grows  through $\gamma_n= \cos (\theta_n/2)$, 
  two opposite eigenvalues $\lambda=-2\alpha$ converge at  $\lambda=0$ and become complex.
  The $\mathcal{PT}$-symmetry breaking threshold is
 determined by the $\gamma$-intercept of the smallest circle: $\gamma_c=  \cos (\theta_N/2)$, i.e., 
 \be
 \gamma_c=  \sin \frac{\pi}{2(2N+1)}.
 \label{alt_open}
 \ee

 \section{Periodic alternating chain}
 
Linearising the  periodic  alternating necklace \eqref{Y1}, 
the corresponding secular equation is   ${\tilde {\mathcal D}_{2N}} (\gamma,\alpha)=0$,
where 
\[
{\tilde{ \mathcal D}_{2N}}=\mathrm{det} \, ({\tilde {\mathcal L}}+ 2 \alpha I).
\]
 The matrix ${\tilde {\mathcal L}}$
is given by the same expression \eqref{Y11}, 
with the same $\Gamma_n= (-1)^{n+1} \gamma$, but with $\delta$ replaced with the {\it cyclic} Kronecker symbol $\delta^{(2N)}$:
\[
\delta^{(2N)}_n= \left\{ 
\begin{array}{ll} 
1, & \mbox{if} \ n \ \mbox{mod} \ 2N=0;  \\
0  &  \mbox{otherwise}.
\end{array}
\right.
\]
The determinant ${\tilde {\mathcal D}}_{2N}$ can be expressed via
the ``nonperiodic" determinants \eqref{D}:
\[
{\tilde {\mathcal D}}_{2N}= \mathcal{D}_{2N}- \mathcal{D}_{2(N-1)} -2.
\]
Using \eqref{Y4} the determinant in question is evaluated to be
\[
{\tilde {\mathcal D}}_{2N}= -4 \sin^2 (N \theta /2 ),
\]
with the double roots $ \theta_n= \frac{2n}{N} \pi$, $n=1,2, ..., N$.

As in the open-necklace case,  equations 
\[
\gamma^2+
\alpha^2= \cos^2 (\theta_n/2)
\]
 describe 
concentric circles 
 on the $( \gamma,  \alpha)$-plane.  When $N$ is even, the smallest circle
 corresponds to $n=N/2$ and has zero radius. When $N$ is odd, the smallest circle corresponds 
 to $n=\frac12 (N-1)$; the radius in this case is  $\sin [\pi/(2N)]$. Thus,
 \be
 \gamma_c= \left\{
 \begin{array}{cr}
 0,  & N=\mbox{even}; \\
 \sin \left( \frac{\pi}{2N} \right), & N=\mbox{odd}.
\end{array}
\right.
\label{alt_closed}
\ee

\begin{figure}[t]
 \begin{center}
\includegraphics*[height=35mm,width=\linewidth]{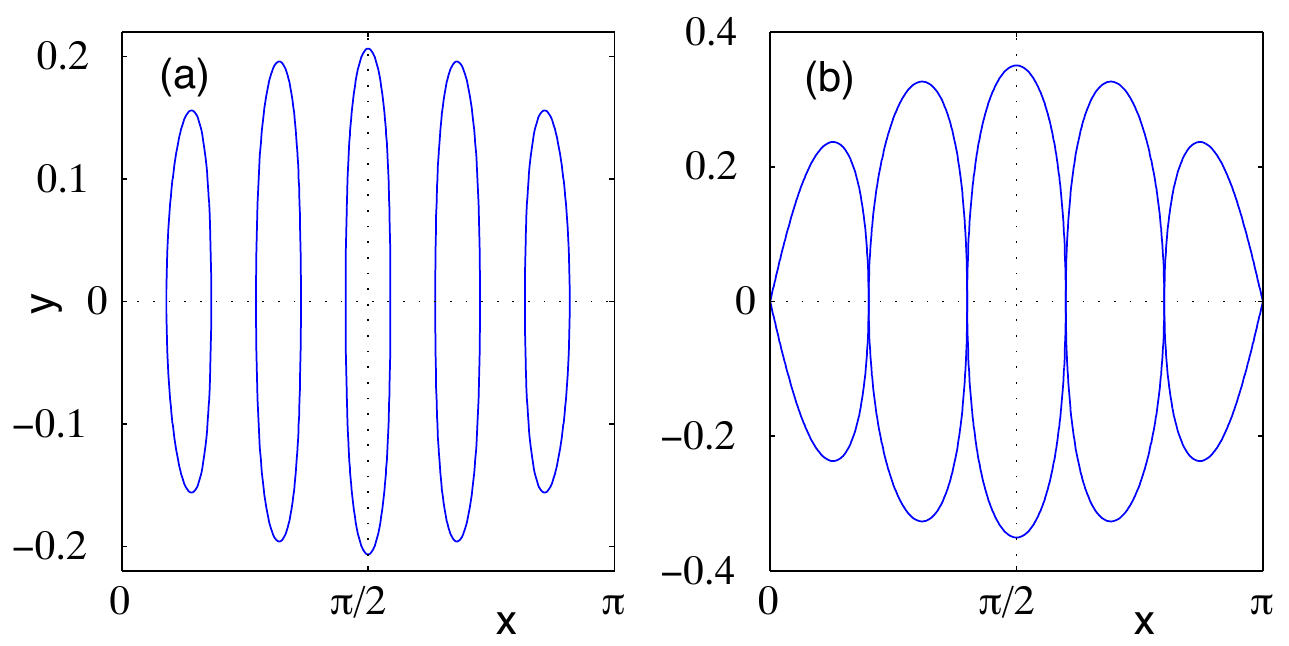} 
            \caption{(Color online) The implicit curves \eqref{B5} (a) and  \eqref{Z3} (b),          in the interval $0 \leq x \leq \pi$.     
            In both panels, $N=5$.                            
                 \label{four_curves}}
 \end{center}
 \end{figure}

\section{Open clustered necklace}

When the waveguides are grouped into two clusters, 
 the gain-loss coefficient $\Gamma_n$ equals $\gamma>0$ for
$n=1, ..., N$ and $-\gamma$ for $n=N+1,  ..., 2N$.
Again, we start with the open chain, Eq.\eqref{Y2}. 
The corresponding linearisation matrix $\mathcal{L}$ is as in Eq.\eqref{Y11}.

To find roots of the corresponding secular equation $\Delta_{2N}=0$, 
where $\Delta_{2N}=\mathrm{det} (\mathcal{L}+ 2 \alpha I)$,
we expand 
 \be
\Delta_{2N}(\gamma, \alpha)= U_N(\zeta) U_N^*(\zeta) -U_{N-1}(\zeta) U_{N-1}^*(\zeta),
 \label{Z1}
 \ee
where  $U_N$ is a determinant of  an $N \times N$ tridiagonal matrix
\[
\mathcal{U}_{m,n}= 2 \zeta \delta_{m-n}+ \delta_{|m-n|-1} \quad  (m,n=1,2..., N),
\]
and $\zeta=  \alpha- i \gamma$. 
This determinant is nothing but the Chebyshev polynomial (of the complex argument);
hence our choice of notation \cite{Cheb}.
Defining complex $\theta$, such that $\zeta= \cos \theta$, the Chebyshev polynomial
can be written as 
\[
U_N(\zeta)= \frac{ \sin[(N+1) \theta]} {\sin \theta}.
\]
Letting $\theta=x +i y$, 
the secular equation reduces to
  \be
\sinh y \sinh [(2N+1) y] = - \sin x \sin [(2N+1) x].
\label{B5}
\ee  
Here $x^2+y^2 \neq 0$ (for $\theta=0$ is not a root of $\Delta_{2N}=0$).  Note that 
$x$ and $y$ are the
  elliptic
coordinates on the $(\gamma,  \alpha)$ plane:
 \be
 \alpha= \cos x \cosh y, \quad \gamma= \sin x \sinh y.
 \label{B10} 
 \ee

 \begin{figure}[t]
 \begin{center}
                     \includegraphics[width=\linewidth]{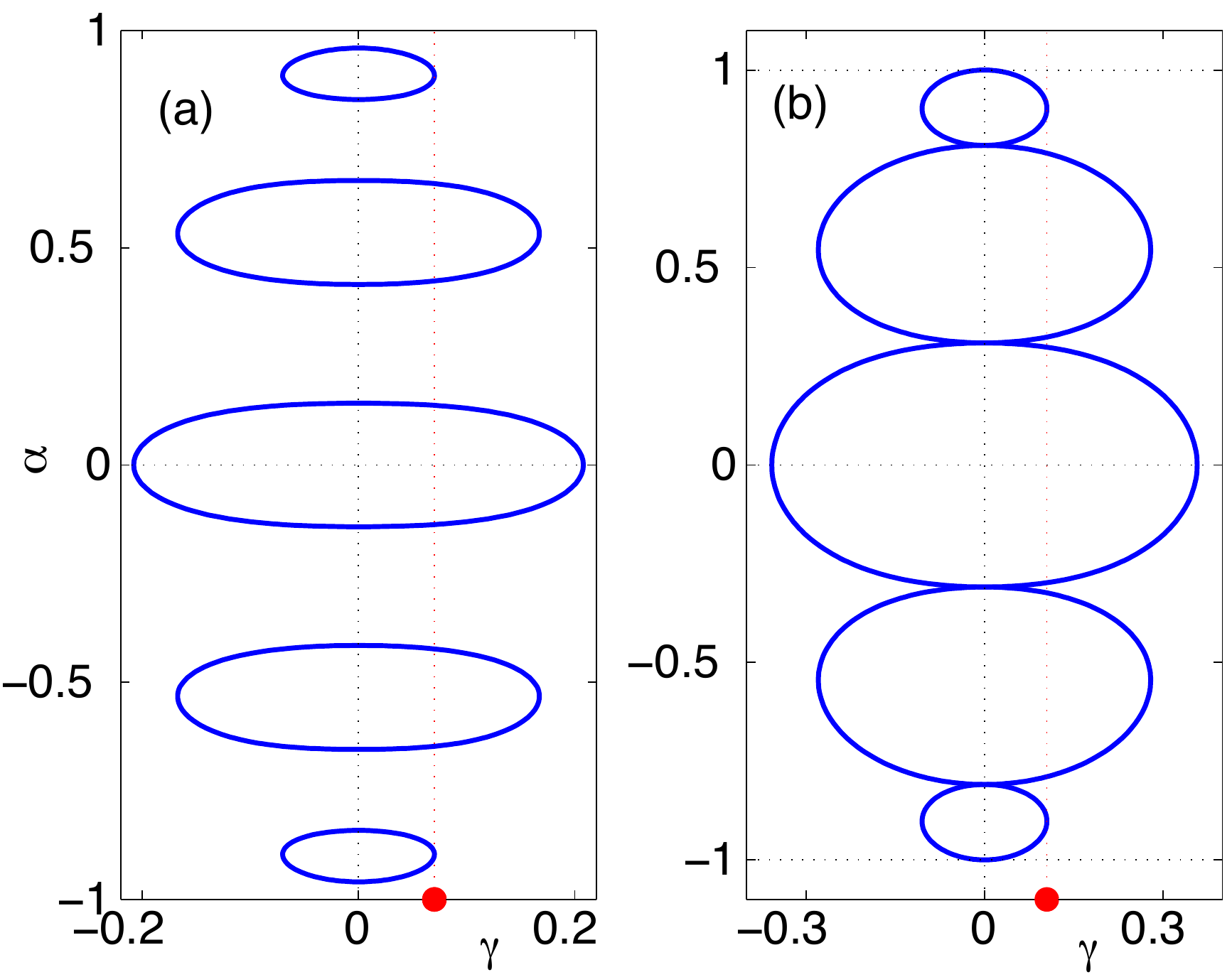}                                      
                  \caption{(Color online) The curve $\Delta_{2N}(\alpha, \gamma)=0$ (a) and ${\tilde \Delta}_{2N}  (\alpha, \gamma)=0$
                  (b) with $\Delta_{2N}$ and ${\tilde \Delta}_{2N}$ as in \eqref{Z1} and \eqref{Z2}. (Here $N=5$.) The red dot on the 
                  $\gamma$-axis
                  marks $\gamma_c$, the point  of the $\mathcal{PT}$-symmetry breaking. 
                     \label{gamma_alpha}}
 \end{center}
 \end{figure}

The  right-hand side of  
\eqref{B5} is  $\pi$-periodic;
hence it is sufficient to consider  the interval $0 \leq x \leq \pi$.
The curve described by \eqref{B5} 
 consists of 
$N$ disconnected ovals in subintervals
\[
\frac{2n-1}{2N+1} \pi \leq x \leq \frac{2n}{2N+1}\pi, \quad
n=1, ..., N
\]
[fig.\ref{four_curves}(a)].
The transformation \eqref{B10} maps these to $N$ ovals on the 
$(\gamma, \alpha)$-plane [shown in fig.\ref{gamma_alpha}(a)]. 
Of interest to us are the points where pairs of $\alpha(\gamma)$ branches merge.

 In Eq.\eqref{B5},  the sinusoide
$\sin (2N+1)x$  is modulated by a slowly changing amplitude $\sin x$. 
Therefore,  of all $N$ ovals, the first and the last one
(those with  $n=1$ and $n=N$) have 
the lowest maximum values of $y$.
The transformation $(x,y)  \to (x, \gamma)$, where  $\gamma= \sin x \sinh y $,
 keeps the pattern horizontally periodic
but  elongates the central ovals still further.
Therefore, 
the lowest value of $\gamma$ for which the merger of two real eigenvalues $\lambda=-2\alpha(\gamma)$ occurs, 
corresponds to the apogee
of the first and the last $\gamma(x)$ ovals.  Using  \eqref{B5},
  the condition 
   $ d \gamma/ dx=0$
    translates into
  \be
\frac{\tanh y }{\tan x} =
 \frac{   [\sin x \sin (2N+1) x]_x} { [ \sinh y \sinh (2N+1) y] _y}.
 \label{B7}
\ee

One can readily 
construct asymptotic roots of the system \eqref{B5}, \eqref{B7}, as $N \to \infty$.
Expanding $x$ and $y$  in powers of $\frac{1}{2N+1}$,
Eqs.\eqref{B5} and \eqref{B7}  give, respectively:
\begin{align}
\xi \sinh \xi  =   -   \mathcal{S} \eta \sin \eta,
\label{B6} 
\\
\xi (\sinh \xi + \xi \cosh \xi)
= \mathcal{S}
\eta (\sin \eta + \eta \cos \eta),
\label{B8}
\end{align}
where
\[
x=\frac{\eta}{2N+1}+ ..., 
\quad
y=\frac{\xi}{2N+1}+ ..., 
\]
and $\mathcal{S}=1$.  

\begin{figure}[t]
 \begin{center}
  \includegraphics[width=\linewidth]{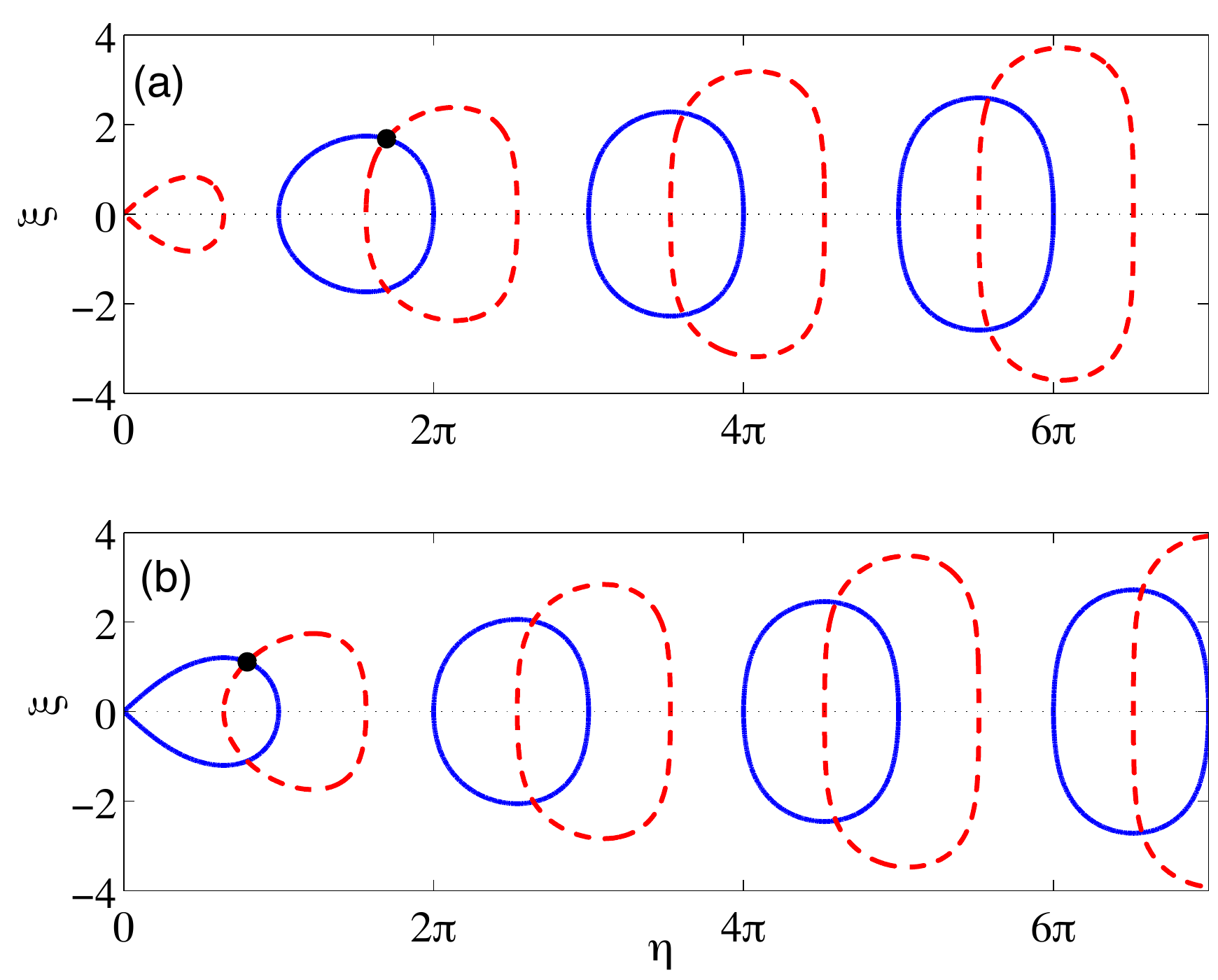}
                 \caption{(Color online) The graphical solution of the system
            \eqref{B6}-\eqref{B8} with $\mathcal{S}=1$ (a) 
            and $\mathcal{S}=-1$ (b). The blue (solid)  and red (broken)  curve are described by equation \eqref{B6}
            and  \eqref{B8}, respectively.
            The black dot marks  the root $(\eta_1, \xi_1)$ in (a),
            and the root $({\tilde \eta}_1, {\tilde \xi}_1)$ in (b). 
                \label{graphical_roots}}
 \end{center}
 \end{figure}

The system \eqref{B6}-\eqref{B8} has an increasing
sequence of roots  $\eta_n, \xi_n>0$, $n=1,2,...$.
[See fig.\ref{graphical_roots}(a).]
In particular, $\eta_1= 5.33$, $\xi_1= 1.68$.
Hence we get, for each $N$,
\[
x_n^{(N)}= \frac{\eta_n}{2N+1}+..., \quad
y_n^{(N)}= \frac{\xi_n}{2N+1}+...,
\]
where ``..." stand for corrections of order $(2N+1)^{-2}$.
(These asymptotic expressions are accurate  for $n$  such that  $\eta_n$
and $\xi_n$ are much smaller than $2N+1$.)

The corresponding values $\gamma_n$ are recovered from \eqref{B10}.
Of primary importance is the smallest value $\gamma_1$
which determines the symmetry breaking threshold:
$\gamma_c=  \gamma_1 =
\eta_1 \xi_1 (2N+1)^{-2}+ ...$. Substituting for $\eta_1$ and $\xi_1$,  we get
\be
\gamma_c
  = \frac{8.95}{(2N+1)^2} +  O \left( \frac{1}{(2N+1)^3}\right)
    \quad \mbox{as} \ N \to \infty.
    \label{oc}
     \ee

\section{Periodic clustered necklace}

The closed clustered chain is described by Eq.\eqref{Y1} with 
$\Gamma_n= \gamma>0$ ($n=1,2, ..., N$) and $\Gamma_n=-\gamma$ ($n=N+1,  ..., 2N$).
The $\mathcal{L}$-matrix is
\[
\mathcal{L}_{nm}= - 2 i \Gamma_n
\delta^{(2N)}_{n-m} + \delta^{(2N)}_{n-m -1}+ \delta^{(2N)}_{n-m+1}, 
\]
$n,m=1, ..., 2N$.

The characteristic determinant
is expandable as 
\be
{\tilde \Delta}_{2N} = U_N U_N^*-2U_{N-1}U_{N-1}^* + U_{N-2}U_{N-2}^*-2,
 \label{Z2}
 \ee
where
$U_N(\zeta)$  is the Chebyshev polynomial of $\zeta= \alpha -i \gamma$.
Using the elliptic coordinates \eqref{B10}, 
 ${\tilde \Delta}_{2N}=0$  reduces to 
\be
\sinh (Ny) \sinh y= | \sin (Nx) \sin x|.
\label{Z3}
\ee

Because of the periodicity  of the right-hand side of \eqref{Z3}, 
it is sufficient to consider the interval $0 \leq x < \pi$.
The curve \eqref{Z3} consists of  $N$ ovals [fig.\ref{four_curves}(b)]: 
one in each of the 
 subintervals 
 \[
 \frac{2(n-1)}{N}\pi \leq x \leq \frac{2n-1}{N} \pi, 
 \quad
 n=1, ..., \left[\frac{N+1}{2}\right]
 \]
 and one in each of the 
 complementary subintervals
 \[
 \frac{2n-1}{N}\pi \leq x \leq \frac{2n}{N} \pi, 
 \quad
 n=1, ... \left[\frac{N}{2}\right].
 \]
 (Here $[p]$ indicates the integer part of $p$.)
 The former set will be referred to as {\it even\/} ovals, and the latter one as {\it odd}.

When $N$ is large, one can find 
the maximum value  of  the function $\gamma(x)$ in
each subinterval as an expansion in powers of $\frac1N$.
The lowest maximum $\gamma_1$ in the {\it odd\/} set  is
given by
$\eta_1 \xi_1 N^{-2}+O(N^{-3})$,
   where $\eta_1, \xi_1$ is the first pair of roots of the system \eqref{B6}-\eqref{B8}
   with $\mathcal{S}=1$. 
  Substituting their numerical values, we obtain $\gamma_1 =   8.95    N^{-2}+...$.

   The lowest maximum in the {\it even\/} set  is 
given by ${\tilde \gamma}_1 =  {\tilde \eta_1} {\tilde  \xi_1}N^{-2} +...$, 
   where ${\tilde \eta_1}=2.50, {\tilde \xi_1}=1.11$ is the first pair of roots of the system \eqref{B6}-\eqref{B8}
   with $\mathcal{S}=-1$. The maximum ${\tilde \gamma}_1$
 is lower than $\gamma_1$; hence
 it is ${\tilde \gamma}_1$ that determines 
  the 
 symmetry-breaking threshold of the closed clustered necklace.
Substituting for ${\tilde \eta}_1$, ${\tilde \xi}_1$, we finally get
\be
\gamma_c
  = 2.77 \, N^{-2} + O\left( N^{-3} \right)
  \quad \mbox{as} \ N \to \infty.
  \label{X4}
   \ee

 \section{Solitons}

 Assume the $\mathcal{PT}$ symmetry is unbroken, $\gamma < \gamma_c$.  Let ${\vec \phi_n}$ be an eigenvector of  $\mathcal{L}$ 
 pertaining to the (real) eigenvalue $\lambda_n$. 
 The transformation $u_n=\sum \Phi_{nm} \psi_m$, with
 $\Phi=\{ {\vec \phi_1}, {\vec \phi_2}, ..., {\vec \phi_{2N}} \}$,
 casts Eq.\eqref{Y100} in the form
 \be
i {\dot \psi_\ell}+ \psi_\ell^{\prime \prime} +\lambda_\ell \psi_\ell+ 2\sum_{n,m} (\Phi^{-1})_{\ell n} \, \mathcal{N}_n \Phi_{nm} \psi_m=0,
 \label{Y200}
 \ee 
 where 
 \[
 \mathcal{N}_n=|\sum \Phi_{nm}\psi_m|^2.
 \]
 (Physically, $\mathcal{N}_n=|u_n|^2$  has the meaning of the power density in the $n$-th waveguide.)
 We will show that the vector
 equation \eqref{Y200}
 admits $2N$ independent scalar  reductions.

 In this paper, we confine our consideration to the case of the periodic alternating chain, 
 Eq.\eqref{Y1} with $\Gamma_n=(-1)^{n+1}\gamma$. The corresponding
  matrix $\mathcal{L}$ has 2 simple and $N-1$ double eigenvalues. The 
 simple eigenvalues $\lambda_{\pm}=\pm 2 \cos \vartheta$ 
 have eigenvectors \[
 {\vec \phi_\pm}=\left(1, \pm e^{\pm i\vartheta}, 1, \pm e^{\pm i \vartheta},...\right),
 \]
 where $\sin \vartheta= \gamma$. These satisfy $| ({\vec \phi_\pm} \,)_j|=1$,
 for all $j$. 
  We now show that the rest of the eigenvectors can also be chosen to satisfy this property.

 The matrix $\mathcal{L}$ commutes with the $\mathbb{Z}_N$-rotation $\mathcal{R}$, where
 $\mathcal{R}_{nm}=\delta^{(2N)}_{n-m-2}$. Therefore the basis in 
 the invariant subspace $S_n$ associated
 with the eigenvalue $\lambda_n$ can be chosen in the form of two eigenvectors of $\mathcal{R}$:
 \be
 \mathcal{R} {\vec \psi_1}= e^{2 \pi i/N}{\vec \psi_1},
 \quad 
 \mathcal{R} {\vec \psi_2}= e^{-2 \pi i/N}{\vec \psi_2}.
 \label{Y300}
 \ee
 The linearisation matrix
  satisfies $\mathcal{L} {\mathcal P}=
 \mathcal{P}  \mathcal{L}^*$, where $\mathcal{P}$ is the inversion: 
\[
\mathcal{P}_{nm}=\delta^{(2N)}_{n+m-1}.
\]
  Therefore, $\mathcal{P} {\vec \psi_1}^*$
and $\mathcal{P} {\vec \psi_2}^*$ are also in $S_n$. Since $\mathcal{R} \mathcal{P}=
\mathcal{P} {\mathcal R}^{-1}$, the vector
$\mathcal{P} {\vec \psi_1}^*$ is an eigenvector of the rotation $\mathcal{R}$, with an eigenvalue $e^{2 \pi i/N}$. That is, 
$
\mathcal{P} {\vec \psi_1}^*= C {\vec \psi_1}$,
 with some constant $C$. Since $\mathcal{P}^2=I$, 
the constant $C=e^{i \chi}$,  with $\chi$         real. Thus,
\be 
\mathcal{P} {\vec \psi_1}^*= e^{i \chi}{\vec \psi_1},
\quad
 \mathcal{P} {\vec \psi_2}^*= e^{-i \chi}{\vec \psi_2}.
 \label{Y400}
 \ee

We normalise ${\vec \psi_1}, {\vec \psi_2}$ so that $({\vec \psi_1})_1= ({\vec \psi_2})_1 =1$. 
Eq.\eqref{Y300} tells us that 
\[
({\vec \psi_1})_{1+2\ell}=e^{-2\pi i \ell/N},
\quad
  ({\vec \psi_2})_{1+2\ell}=e^{2\pi i \ell/N}, 
  \quad
  \ell=1,2,... .
  \]
On the other hand, Eq.\eqref{Y400} gives 
\[
({\vec \psi_1})_{2N-2 \ell}= e^{-i \chi+ 2 \pi i \ell/N},
\quad
 ({\vec \psi_2})_{2N-2 \ell}= e^{i \chi- 2 \pi i \ell/N}.
 \]
Thus all eigenvectors  of $\mathcal{L}$
have unimodular components:
\be
|\Phi_{nm}|= |({\vec \phi_m})_n|= 1;  \quad n,m=1,2,..., 2N.
\label{Y700}
\ee

Returning to \eqref{Y200}, the scalar reduction is defined by letting
$\psi_{m}= \psi \delta_{m -M}$, with some fixed $M$. In 
view of \eqref{Y700}, this gives $\mathcal{N}_n=  |\psi|^2$ for all $n$.
All components of \eqref{Y200} become identically zero, except the one with $\ell=M$,
which becomes
\be
i {\dot \psi}+ \psi^{\prime \prime} + \lambda_M \psi+ 2 |\psi|^2 \psi=0.
 \label{Y800}
 \ee 
Each nonlinear Schr\"odinger equation \eqref{Y800}, with $M=1,...,2N$, 
supports a soliton \[
\psi=e^{i \Omega t} a \, \mathrm{sech\/} (az), 
\]
with the frequency
$\Omega=a^2+ \lambda_M$. Thus the original $\mathcal{PT}$-symmetric system
\eqref{Y1} has $2N$ coexisting soliton solutions, different in their frequencies and
polarisations.

We should  emphase the difference between these vector solitons 
and (spatial) solitons in a $\mathcal{PT}$-symmetric optical lattice
  \cite{Musslimani2}. 
While
 the solitons  in
the waveguide necklace \eqref{Y1} are localised 
as functions of $z$,
the lattice solitons \cite{Musslimani2} are localised as functions of $n$
(i.e., in the transverse direction).
The  $n$ dependence 
determines the  power density distribution over the $2N$ channels;
this distribution is uniform in the case of the vector solitons
of Eq.\eqref{Y1}.
  
  \section{Concluding remarks}
  
In conclusion, we have determined the symmetry breaking  points 
for four different geometries  of the necklace.
Generically, there is a finite interval of the gain-loss coefficient  where the $\mathcal{PT}$ symmetry is unbroken.
The  only exception is the periodic chain of $2N$ alternating waveguides
with even $N$. Here $\gamma_c=0$, i.e., the  symmetry is spontaneously broken  for an arbitrarily small $\gamma$.

The {\it alternating\/} arrays admit an explicit solution; the transition points are given 
by \eqref{alt_open} for the open chains and by \eqref{alt_closed} for the periodic necklaces.
In both cases $\gamma_c \sim \frac1N$ for large $N$. 
In the {\it clustered\/} geometry, the threshold values are expressible 
via roots of 
a simple transcendental equation --- Eq.\eqref{B5} in the case of the open 
chain, and Eq.\eqref{Z3} in the periodic situation. 
Eqs.\eqref{oc} and \eqref{X4} yield the corresponding asymptotic results.
 Here, the $\mathcal{PT}$ symmetry breaks quicker:
$\gamma_c \sim \frac{1}{N^2}$ as $N \to \infty$.

It is interesting to note that a similar $\gamma_c \sim \frac{1}{N^2}$ law was detected
 in a {\it disordered\/} $\mathcal{PT}$-symmetric chain with the clustered 
arrangement of gain and loss, in the limit of large localisation lengths of the eigenmodes
\cite{Bendix}.

Our  $\gamma_c(N)$ values remain valid for the 
arrays of {\it nondispersive\/} $\mathcal{PT}$-symmetric couplers [Eqs.\eqref{Y1}-\eqref{Y2}
without the $u_n^{\prime \prime}$ term]. In particular, our conclusion  that the 
limit of the sequence $\gamma_c(N)$ as $N \to \infty$ exists and equals 0, 
is in agreement with the symmetry-breaking threshold for the infinite alternating  chain \cite{Dmitriev}.

Finally, we have demonstrated that the alternating periodic necklace supports $2N$
coexisting soliton solutions. These vector solitons are characterised by the uniform distribution
of the power density over their $2N$  components and are different in their frequencies and polarisations.
It is natural to expect that the other three $\mathcal{PT}$-symmetric waveguide arrangements
(open-alternating,  open- and periodic-clustered) will
also exhibit $2N$ different solitons each.

\end{document}